\documentclass[%
 aip,
 rsi,
 amsmath,amssymb,
 reprint,
]{revtex4-2}

\usepackage{graphicx}
\usepackage{dcolumn}
\usepackage{bm}

\usepackage[utf8]{inputenc}
\usepackage[T1]{fontenc}
\usepackage{mathptmx}
\usepackage{etoolbox}
\usepackage[above]{placeins}
\usepackage{siunitx}

\makeatletter
\def\@email#1#2{%
 \endgroup
 \patchcmd{\titleblock@produce}
  {\frontmatter@RRAPformat}
  {\frontmatter@RRAPformat{\produce@RRAP{*#1\href{mailto:#2}{#2}}}\frontmatter@RRAPformat}
  {}{}
}%
\makeatother
\begin{document}

\title[Thomson Scattering Measurement of 1 keV Electron Temperature on Lawson Machine 26]{Thomson Scattering Measurement of 1 keV Electron Temperature on Lawson Machine 26}
\author{P. Carle}
 \email{patrick.carle@generalfusion.com}
\author{S. Edwards}
\author{X. Feng}
\author{A. Froese}
\author{M. Greenwood}
\author{A. Lee}
\author{M. Reynolds}
\author{A. Rudy}
\author{J. Sanchez Rojo}
\author{M. Schellenberg-Beaver}
\author{R. Svihra}
\author{R. Tingley}
\author{R. Zindler}
\affiliation{General Fusion Inc., Vancouver, British Columbia, Canada}
\author{M. Maslov}
\affiliation{UKAEA, Culham Campus, Abingdon, Oxfordshire, OX14 3DB, UK}

\date{\today}

\begin{abstract}

LM26 is a magnetized target fusion experiment built to demonstrate compressional heating of a plasma with a collapsing solid lithium liner. We report the design of a Thomson scattering diagnostic for LM26 and measurements from the LMC-12 compression shot where an electron temperature of $1090\pm40~\rm{eV}$ was measured with high confidence. The LM26 Thomson scattering diagnostic can measure temperature during compression at 8 times and 2 spatial points. Access limitations restricted the design to narrow f/30 views, and a forward scattering configuration with scattering angles $<20\deg$. The small solid angle of the scattered light view is largely compensated for by the 10 times increase in the density during compression to $>10^{20}\rm{m}^{-3}$. Measuring at a small scattering angle significantly narrows the Thomson spectral broadening of the laser line. The narrow spectrum (1030-1063~nm) is advantageous for the design of optical components and to minimize noise from plasma background light. UKAEA polychromators specialized for small broadening measurements were used to analyze the scattered light. Results are also presented for an uncompressed plasma preceding LMC-12. Stray light was not a significant source of error. This design provides a strong design basis for Thomson scattering systems on future LM26 configurations. 

\end{abstract}

\maketitle

\section{Introduction}
\begin{figure}
    \centering
    \includegraphics[width=1\linewidth]{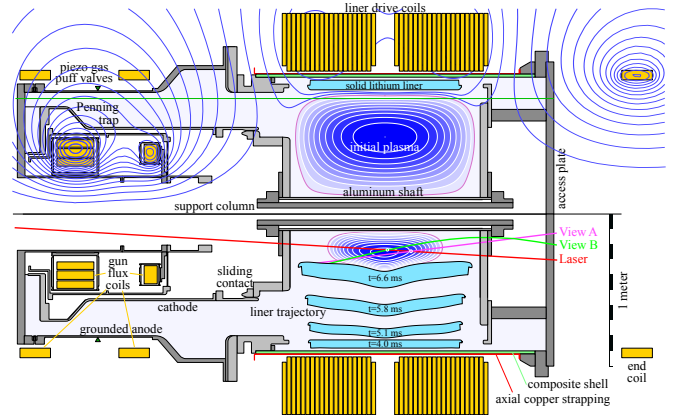}
    \caption{LM26 cross section with Thomson laser and view chords for the LMC-12 compression shot. Poloidal flux contours are shown in the target chamber for an initial plasma and a compressed plasma. Liner positions are shown at several times during compression with $t=0$ and $t=4.0~\rm{ms}$ respectively being the times of plasma formation and the start of compression. The Thomson laser and view chords are projections of the 3-dimensional chord onto $(R,Z)$. The View B chord appears curved because it starts and ends at different toroidal angles.}
    \label{fig:lm26}
\end{figure}

LM26 is a magnetized target fusion (MTF) experiment designed to demonstrate heating of a magnetized deuterium plasma by compressing it with an imploding solid lithium liner \cite{howard2026science}. A cross section of LM26 is shown in Figure~\ref{fig:lm26}. A Marshall gun injects a plasma into a cylindrical target chamber. The target chamber is a composite (non-conducting) vacuum vessel with a solid lithium cylinder inserted on the inside. Current is driven through the aluminum central shaft of the machine to generate toroidal flux to improve the magnetic stability of the plasma. The initial plasma has electron temperature $T_e\approx200~\rm{eV}$, electron density $n_e\approx3 \times 10^{19}~\rm{m}^{-3}$ and magnetic field $B\approx0.5~\rm{T}$. The plasma settles in the target chamber for 2-4 ms before compression is initiated by pulsing current through the liner drive coils located on the outside of the vacuum vessel. The generated magnetic field pushes on the lithium liner such that the liner collapses around the plasma and compresses the plasma to a smaller volume. The time from start of compression to peak compression is about 3~ms. If the plasma were not compressed, it would resistively decay for around 20~ms. The present configuration of LM26 is designed to compress plasmas to 1~keV with a liner radial compression ratio of $C_R\approx3$. With planned upgrades to the machine, LM26 is designed to next demonstrate plasma compression to 10~keV.

The compression shot cycle for a newly assembled machine begins with a series of plasma shots to improve the wall conditions and optimize plasma performance. Lithium coaters are inserted into the machine to evaporatively coat the plasma facing surfaces with a micron-thick layer of lithium, which is a good getter and has been shown to significantly improve plasma performance \cite{Pi3_tauE_NucFus_2025}. When the compression shot is completed, diagnostics are removed, and the shaft and  spent liner are extracted by removing an access plate. A new shaft and liner are then inserted under argon atmosphere to protect machine components from contamination. Once the machine is assembled, it is once again pumped to high vacuum. Diagnostics are reinstalled and the shot cycle for the next compression campaign begins.

The presence of the liner and Marshall gun constrains most plasma diagnostic chords to pass through ports on the access plate. When the compression coils are fired, diagnostics must be able to withstand the ensuing electromagnetic interference, vibration and mechanical shock. In addition, diagnostics must be designed to be quickly uninstalled, re-installed and calibrated in between compression shots to avoid delaying the shot schedule.

Measurement of the electron temperature, $T_e$, is an important indicator of compression performance and has been primarily measured with a multi-energy soft X-ray diagnostic until a single-view Thomson measurement was added for the LMC-11 compression shot \cite{howard2026science}. This paper presents the design of the LM26 Thomson system and recent results from LMC-12 which includes a 1~keV electron temperature measurement.

\section{Thomson Scattering on LM26}

The following sections describe the design of the Thomson scattering diagnostic for LM26. While a prototype of the Thomson system was tested for LMC-11 and did generate data \cite{howard2026science}, this paper focuses on the LMC-12 Thomson scattering diagnostic design which had additional features of interest.

\subsection{Laser Path}

The LM26 Thomson system uses two Nd:YAG lasers (InnoLas SL2500-HP-1064-10Hz) which can individually produce 4-pulse bursts at a 250~\unit{\micro\second} interval and with 1.9~J energy in each 10~ns pulse at a wavelength of 1064~nm. The beams are combined with a polarizer (Thorlabs GL15-C26) by rotating the polarization of one of the beams with a half-wave plate so that the beam polarizations are orthogonal to one another. The combined beams are collinear, making it possible to double the number of pulses from 4 to 8 and to double the pulse frequency by interleaving laser pulses. A second half-wave plate is placed after the polarizer to be able to rotate the polarization of the combined beams.

\begin{figure}
    \centering
    \includegraphics[width=.80\linewidth]{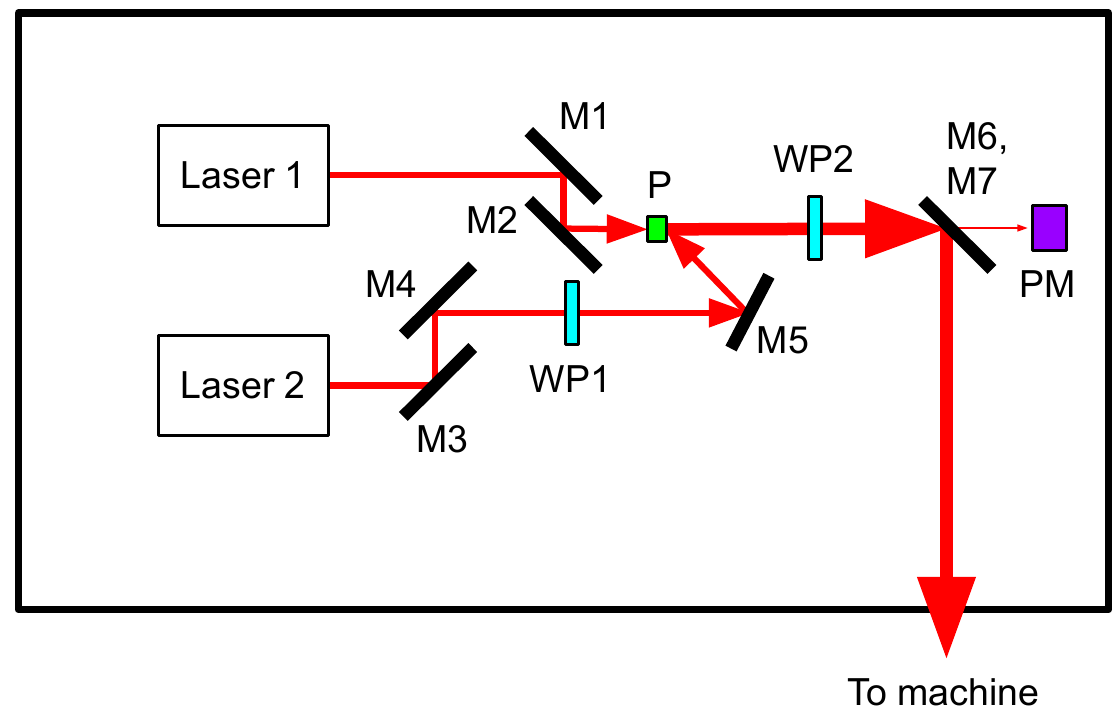}
    \caption{Diagram of Thomson laser table optical layout. The Laser 2 beam polarization is rotated 90 degrees by a half-wave plate, WP1, to combine the beam with the Laser 1 beam at the polarizer, P. The polarization of the combined beams can be rotated again with half-wave plate WP2 to optimize stray light at the Brewster window on the machine. Mirror M6 deflects the beam upwards above head height and mirror M7 steers the beam towards the machine. The laser pulse monitor, PM, collects stray light leaking through M6.}
    \label{fig:laser_table}
\end{figure}

\begin{figure}
    \centering
    \includegraphics[width=1\linewidth]{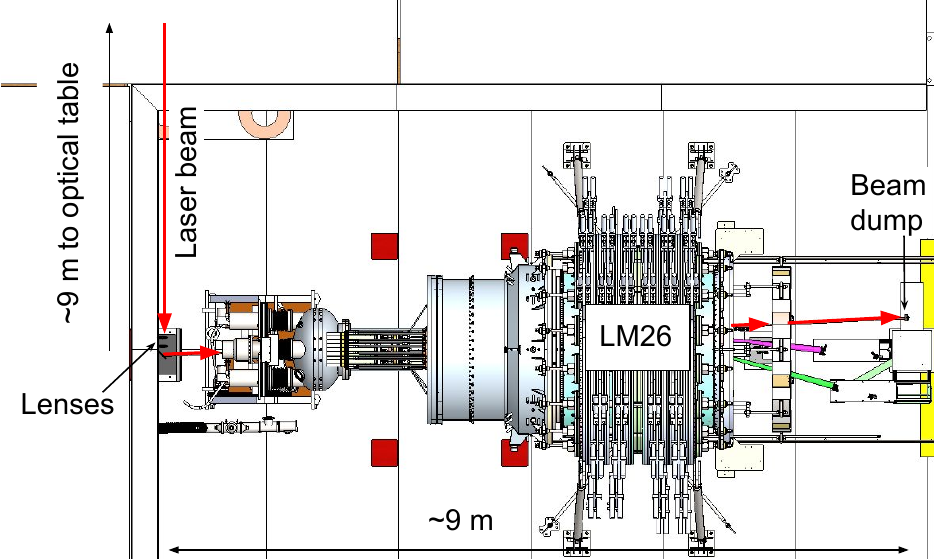}
    \caption{Top view of the Thomson laser beam path through the machine. The laser table is not shown but is about 9~m away from the lenses.}
    \label{fig:beam_path_top}
\end{figure}

Figure~\ref{fig:laser_table} shows a diagram of the optical layout on the laser table and Figure~\ref{fig:beam_path_top} shows the top view  the laser beam's path through the machine. Both lasers are located inside a laser room near the machine and the beam exits the room via an interlocked safety shutter. Before entering the machine, the beam passes through a pair of lenses with focal length $f=+1~\rm{m}$ (Newport KPX217AR.33) and $f=-1~\rm{m}$ (Newport SPC037AR.33)  with the distance between them set to focus the beam at the equatorial plane of the compression chamber.

\begin{figure}
    \centering
    \includegraphics[width=\linewidth]{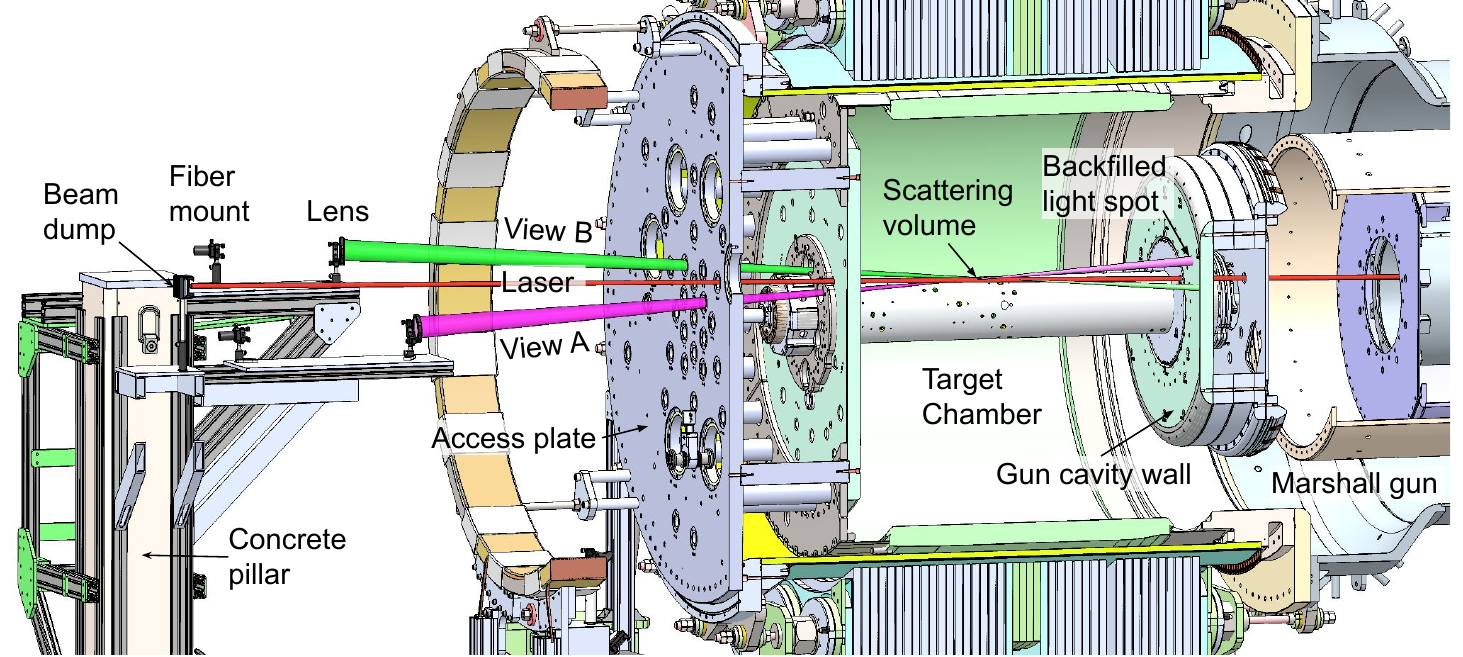}
    \caption{Side view of Thomson scattering geometry for LMC-12. Enclosures for collection optics and laser beam are not shown.}
    \label{fig:beam_path_side}
\end{figure}

The laser beam is steered through the machine from the Marshall gun to the access plate (see Figure~\ref{fig:beam_path_side}). The beam enters the machine through a planar antireflection coated window (Thorlabs VPW42-C) oriented nearly normal to the laser beam propagation direction. The laser beam diameter at this window is about 15~mm, which is sufficiently large to keep the laser energy density well below the window's damage threshold. The beam diameter at the exit window is only 6~mm since the location of the machine's access plate is relatively close to the beam's focal point. Due to the higher laser energy density at this location, the output window is an uncoated Brewster window (Edmund 37-690) with orientation between the s- and p- polarizations of the two lasers. The second half-wave plate on the laser table could have been used to maximize transmission through the Brewster window and minimize stray light, but it was not necessary in the end.

The laser beamline is enclosed in metal pipes for the whole path around the machine. The beam is terminated at a beam dump (Thorlabs LB2) placed 2~m away from the Brewster window to allow the beam to expand before striking the dump to reduce ablation at the dump's surface and associated back-reflected stray light with broadband spectrum. All laser optics other than the vacuum windows are mechanically isolated from the machine in order to minimize vibrations and maximize alignment stability as much as possible. 

\subsection{Scattered Light Collection}

Diagnostic access to the plasma in LM26 is constrained by the surrounding liner and Marshall gun. A more typical scattering angle of 90~deg would require installation of collection optics within the central shaft. While feasible, this option was abandoned due to complications and risks related to machine vibration, protection of optics during lithium coating, and recovery of diagnostic components post-compression. Therefore, the diagnostic design proceeded with a forward scattering configuration at a small scattering angle $\theta<20~\deg$ which has been accomplished previously on other machines \cite{Yamada2025LHD}.

\begin{figure}
    \centering
    \includegraphics[width=\linewidth]{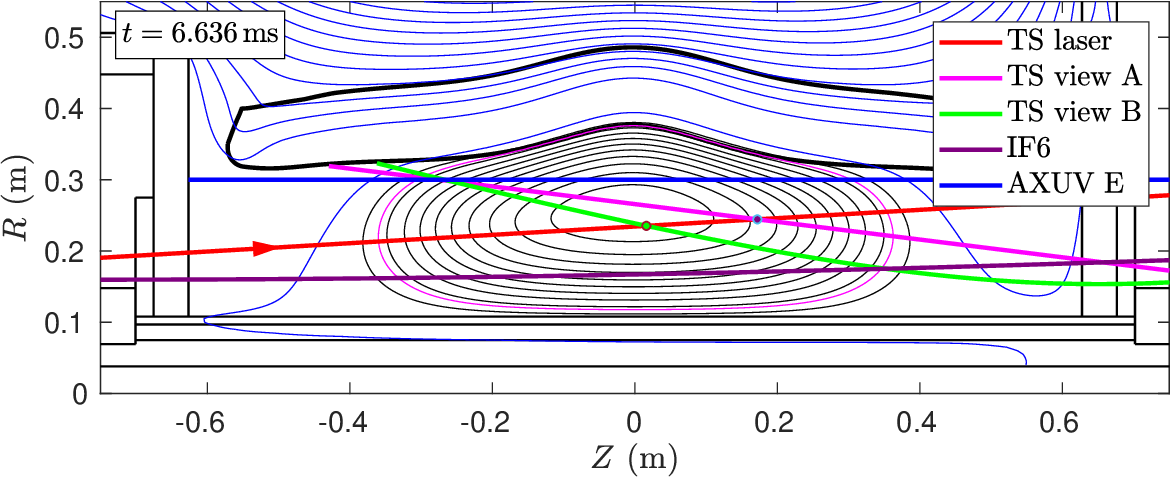}
    \caption{Thomson scattering laser (TS laser) and view lines (TS view A, TS view B), interferometer chord (IF6), and soft X-ray diagnostic view (AXUV E) projected onto $(R,Z)$. Liner and poloidal flux contours shown for compressed plasma in LMC-12 at the time of the 6th Thomson pulse ($t=6.636~\rm{ms}$).}
    \label{fig:lm26_diagnostic_chords}
\end{figure}

\begin{figure}
    \centering
    \includegraphics[width=.8\linewidth]{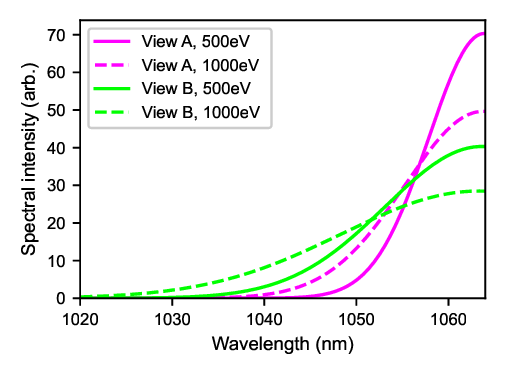}
    \caption{Simulated Thomson broadening for view A and B at 500~eV and 1000~eV. View A and B respectively have scattering angles of $10.4~\deg$ and $18.2~\deg$.}
    \label{fig:expected_broadening}
\end{figure}

As shown in Figure~\ref{fig:beam_path_side} and Figure~\ref{fig:lm26_diagnostic_chords}, two scattered light collection views are used. Table~\ref{tab:view_params} summarizes the viewing parameters. View A intersects the laser beam about 17~cm off the equatorial midplane with scattering angle $\theta_A=10.4\pm0.2~\deg$. View A was previously implemented on LMC-11 \cite{howard2026science}. View B observes the scattered light near the equatorial plane with $\theta_B=18.2\pm0.2~\deg$ scattering angle. See Appendix~\ref{sec:scattering_angle_calculation} for details on determination of the scattering angle. Despite the small scattering angles, the Salpeter parameter is much less than 1 at the expected measurement conditions of $10^{20}~\rm{m}^{-3}$ and 1~keV. Therefore, the Thomson measurement is in the incoherent regime . For a small scattering angle, the expected spectral broadening is relatively narrow at the temperatures of interest as shown in Figure~\ref{fig:expected_broadening}. For a 1~keV measurement, the useful part of the spectrum is in the range of 1030-1063~nm, which is much narrower than in a more conventional 90 degrees scattering configuration. This is an advantage to the design since it minimizes issues with chromatic aberration and spectral transmission of optical components, and also reduces plasma background light contamination of the Thomson signal. Another advantage of a forward scattering configuration is the weak sensitivity of the scattering signal to the laser polarization since $\cos^2\theta$ is close to 1 at small angles.

\begin{table}[htbp]
\centering
\caption{Thomson scattering viewing parameters.}
\label{tab:view_params}
\begin{tabular}{lll}
\hline
\textbf{Parameter} & \textbf{View A} & \textbf{View B} \\
\hline
Scattering angle (deg) & $10.4\pm0.2$ & $18.2\pm0.2$ \\
Scattering length (cm) & $3$ & $3$ \\
Distance from scattering volume to lens (cm) & 194 & 234 \\
Lens focal length (cm) & 50 & 40 \\
Distance from lens to fiber (cm) & 67 & 48 \\
Solid angle (ster.) & $9.5 \times 10^{-4}$ & $7.2 \times 10^{-4}$ \\
\hline
\end{tabular}
\end{table}

The scattering volumes were viewed through broadband antireflection coated vacuum windows (Accu-Glass 115063-BB1) with 35~mm distortion-free view diameters. The light was collected by 75~mm diameter plano-convex lenses (Edmund Optics 86-925 and 71-862). Scattering lengths were approximately 3~cm for both views. For each view, the scattered light was imaged onto a $3\times2~\rm{mm}$ rectangular optical fiber bundle containing 130 fibers and with numerical aperture of 0.285. The fiber was oriented with the long side perpendicular to the laser beam to facilitate the alignment. The views capture light at around f/30, which is relatively poor for a typical Thomson scattering system. However, during compression the density increases by a factor of 10 to several times $10^{20}~\rm{m}^{-3}$, which compensates for the large f/\#.

Alignment of the collection optics is not possible via Raman scattering since the required gas fill of the machine (e.g. nitrogen) would react with and degrade the lithium inside the vacuum vessel. Alignment by Rayleigh scattering with an argon gas fill is challenging due to contamination of the Rayleigh signal with stray light and due to the presence of microdust particles suspended inside the machine when filled with gas. The collection optics were aligned by backfilling the fiber with white light and projecting the light into the Thomson viewing window. The light cone was centered on 2 reference points: the centre of the viewing window and a pre-determined point on the gun cavity wall (see Figure~\ref{fig:beam_path_side}) inside the machine. A camera/re-entrant lens system, which was normally used for imaging the plasma/liner during shooting, was repurposed to view the light spot on the gun cavity wall. For View B, a small hole was drilled into the gun cavity wall before the machine was assembled to serve as a fiducial for the collection optics alignment. It was unfortunately not possible to make a fiducial on the gun cavity wall for View A, and so the light spot position was aligned using nearby port holes as reference points and comparing to the desired spot position from a 3D model of the machine (see Appendix~\ref{sec:scattering_angle_calculation}). The alignment was confirmed by observing Thomson scattering signal during the cleaning/plasma-tuning phase of the campaign leading up to the compression shot.

\subsection{Signal Detection}

\begin{figure}
    \centering
    \includegraphics[width=.7\linewidth]{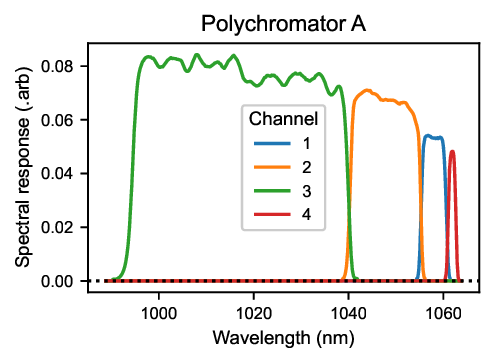}
    \includegraphics[width=.7\linewidth]{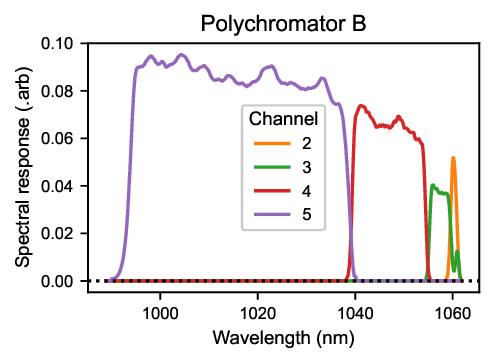}
    \caption{Polychromator A (left) and B (right) spectral calibrations. For details on the calibration process, see Appendix~\ref{sec:poly_calibration}.}
    \label{fig:poly_cals}
\end{figure}

For each view, the collected light is transmitted through a fiber bundle and delivered to a UKAEA polychromator \cite{clark2021} with 4 spectral channels distributed in the wavelength range of 990-1063~nm (see Figure~\ref{fig:poly_cals}). These polychromators are designed for low temperature 1-100~eV measurements at a $90~\deg$ scattering angle but, for the small scattering angle on LM26, they are able to resolve temperatures in the 30~eV to 10~keV range due to much smaller broadening of the Thomson scattering spectrum. The low temperature UKAEA polychromators have at least OD7 (optical density 7) rejection of stray light, but the polychromator for view B was customized with an additional OD2 rejection since the stray light was initially a major concern for these small scattering angle measurements. This resulted in a re-ordering of the spectral channels and a small spectral overlap between channels B2 and B3 which in the end had no effect on the quality of the measurements. In addition to the 4 regular spectral channels, the view B polychromator had a channel (B1) dedicated to monitoring stray laser light at 1064~nm with approximately OD2 suppression (channel B1 is not shown on Figure~\ref{fig:poly_cals}). The stray light was determined to be an insignificant source of error as will be shown in a later section. While the additional stray light rejection features were not needed in the end, channel B1 was a useful monitor of stray light during the compression shot and could be convenient in the future if alignment by Rayleigh scattering is revisited.

For each polychromator spectral channel there are slow and fast outputs both of which are digitized. The slow channels are DC-coupled and monitor background plasma light for the duration of the plasma shot. The fast channels are high pass filtered to remove the relatively slow background light offset and to pass through the fast 10~ns Thomson pulse. For each fast channel, a 500~ns time window is digitized with a 200~MHz bandwidth, 8-bit resolution oscilloscope channel at 1~GS/s. The fast scopes were triggered with a fiber that collected stray laser light leaking through a mirror on the laser table ("PM" in Figure~\ref{fig:laser_table}). The time window was sufficient to record the Thomson scattering signal and the background noise, which is the combination of intrinsic noise of the entire electrical stack (detector, amplifier, digitizer) and the statistical fluctuations (quantum noise) of the plasma background light. The amplitude of the scattered signal is determined by fitting a Gaussian to the measured fast data. 

For an uncompressed plasma density of $10^{19}~\rm{m}^{-3}$, we expected to detect a total of ~250 photoelectron events in each Thomson pulse taking into account the scattering geometry, collection optics transmission losses of around 50\% mostly from coupling the scattered light to the fiber bundle, and an effective quantum efficiency (EQE) of 0.1 for the polychromators' Si detectors. This is sufficient to observe the scattered light in an uncompressed plasma which is helpful to confirm that the collection optics are aligned. During compression, the density will increase to $>10^{20}~\rm{m}^{-3}$ which will increase the number of photoelectrons by a factor of 10 and significantly improve the quality of the measurement. 

\subsection{Measurement Timing}

A cold-fired burst of 4 pulses is initiated from either laser with a single trigger. The control computer sends a trigger signal to a function generator which generates a new pulse with the required laser trigger pulse shape. This pulse is split with one branch sent to trigger the first laser and the other branch sent to the trigger input for the second function generator channel which adds a timing delay and generates a second trigger pulse which is sent to trigger the second laser. The delay of the second trigger can be changed to create sequential or interleaved pulse trains. For the compression shot, the pulses were interleaved to provide a $125~\unit{\micro\second}$ interval between pulses to resolve faster changes in the temperature as the plasma rapidly heats in the last millisecond of compression. 

Timing of the control computer laser trigger is an important consideration for a compression shot. We are interested in measuring the temperature at the deepest compression ratio possible. The liner radial compression ratio, $C_R$, is defined as the ratio of the liner equatorial radius pre-compression to the liner equatorial radius at the time of interest. Before the liner crosses the laser chord, we can measure up to a compression ratio of about $C_R=2.5$. Because the laser is angled through the machine (Figure~\ref{fig:lm26_diagnostic_chords}), a compression ratio of up to $C_R=3.0$ could be measured if the laser were shot into the liner on the access plate side of the target chamber before being completely cut off from the plasma on the Marshall gun side. However, there would be a risk of generating significant broadband stray light from the laser ablating the surface of the liner. Given this risk, it is prudent to plan ahead of the compression shot to fire a pulse as close as possible to cutoff at the access plate side of the target chamber but without striking the liner. As described in Ref.~\onlinecite{howard2026science}, the liner trajectory is decided upon and modeled well in advance of a compression shot to accomplish the goals for a particular LMC campaign. Based on the compression shots completed to date, the model tends to be faster than reality by several tens of microseconds on average. With this in mind for LMC-12, we chose to time laser pulse 7 such that it just intersects the liner according to the liner trajectory model. Pulse 8 would then strike the liner which would give us data on the severity of the stray light.

\subsection{Data analysis}

Plasma electron temperature and density are inferred from the Thomson scattering measurements by finding the minimum of the $\chi^2$ parameter\cite{Maslov2012}:
\begin{equation}
    \chi^2 = \sum_{i=1..,n} w_i \left[ S_i - N \times F_i \right]^2
\end{equation}
where $S_{i}$ is the measured signal in polychromator channel $i$, $N$ is a scaling factor chosen to minimize $\chi^2$ and which could be interpreted as the uncalibrated plasma density, $F_i(T_e)$ is the synthetic signal for channel $i$ for a given electron temperature, $w_i=1/\sigma_{s,i}^2$ is the weighting of each channel, and $\sigma_{s,i}$ is the measurement uncertainty in a particular spectral channel with $\sigma_{s,i}^2 = \sigma_{T,i}^2 + \sigma_{b,i}^2$, where $\sigma_{T,i}$ is the photoelectron statistics of the Thomson signal and $\sigma_{b,i}$ is the background noise which includes the plasma background light quantum noise and the intrinsic electronic noise in the amplifier/digitizer chain. The background noise is measured as the standard deviation of the signal fluctuation in a 200~ns time window prior to the Thomson pulse. Thomson signal photoelectron noise is determined from a conversion of signal amplitude to photoelectron counts which is measured by observing the standard deviation of fluctuations on the fast channels when a broadband DC light source is fed into the polychromator. The conversion factor is found to be at least $1.6\times10^4$~photoelectrons per volt. $\sigma_{s,i}^2$ is used as a weighting factor for a spectral channel in $\chi^2$, but it is also used to determine the statistical error bars for the measured temperature, $\sigma_{T_e, poly}$, via the error propagation calculation described in Ref.~\onlinecite{Maslov2012}.

In the forward scattering geometry the calculated electron temperature is strongly dependent on the scattering angle. Therefore the uncertainty in the scattering angle needs to be taken into account as well. The nominal scattering angle, $\theta_0$, is determined from four $(x,y,z)$ reference points which define the vectors for the laser chord and the collection optics view. The details are described in Appendix~\ref{sec:scattering_angle_calculation}. The uncertainty in the scattering angle, $\sigma_\theta$, is calculated numerically to be $\pm0.2\deg$ for both views from a Monte Carlo simulation which computes the scattering angle for 100,000 variations of the reference points randomly sampled from normally distributed uncertainties in the coordinates. To determine the effect of the scattering angle uncertainty on the temperature, the $T_e$ calculation is repeated with scattering angle $\theta_-:=\theta_0-\sigma_\theta$. The $T_e$ uncertainty contribution due to the scattering angle is then $\sigma_{T_e, \theta} = T_{e,\theta_0}-T_{e,\theta_-}$ where $T_{e,\theta_0}$ and $T_{e,\theta_-}$ are the temperatures calculated respectively with $\theta_0$ and $\theta_-$ as the scattering angle. The smaller scattering angle, $\theta_-$, is chosen because the effect on $T_e$ is stronger at smaller scattering angles, and therefore $\theta_-$ gives a more conservative $\sigma_{T_e,\theta}$.

The total uncertainty in temperature is the quadrature sum of the uncertainty contributions from the polychromator measurements and the scattering angle: $\sigma_{T_e} = \sqrt{ \sigma_{T_e, poly}^2 + \sigma_{T_e, \theta}^2 }$.

\section{Measurement of 1 keV on LMC-12}

\begin{figure}
    \centering
    \includegraphics[width=.8\linewidth]{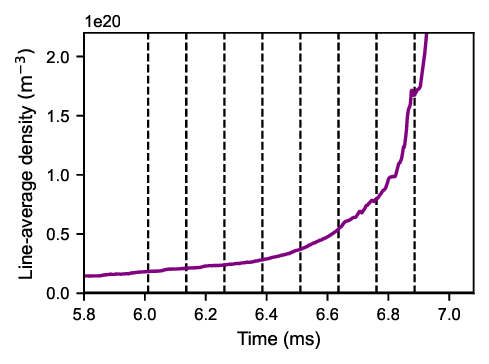}
    \caption{LMC-12 Thomson laser pulse timing as dashed vertical lines. Interferometer line average electron density on chord IF6 is also shown to illustrate the progress of compression. Interferometer IF6 chord length is 1.25~m.}
    \label{fig:laser_timing}
\end{figure}

This section presents Thomson scattering results from the LMC-12 compression shot which resulted in a peak Thomson measurement of $1090\pm40~\rm{eV}$. The timing of the 8 pulses is shown in Figure~\ref{fig:laser_timing}. Compression was initiated at $t=4~\textrm{ms}$ after plasma formation. Thomson laser pulse 1 of 8 occurred at $t=6.01~\textrm{ms}$. Pulse 7 was fired at $t=6.76~\textrm{ms}$, which was 100~\unit{\micro\second} before the liner was in the path of the laser as determined by liner reconstruction \cite{howard2026science}. Pulse 8 struck the liner. The intention of pulse 8 was to measure temperature at a deeper state of compression after the liner had crossed the laser beam path on the access plate side of the target chamber. However, we observed high, broadband stray light levels for pulse 8 and the data was not easily usable.

\begin{figure}
    \centering
    \includegraphics[width=.65\linewidth]{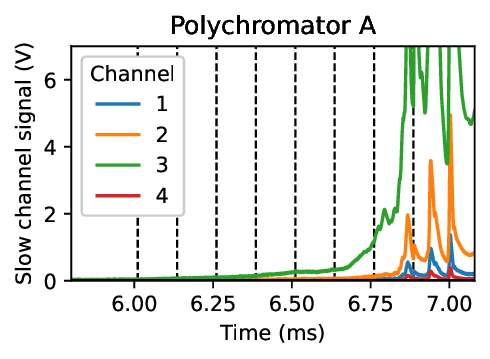}
    \includegraphics[width=.65\linewidth]{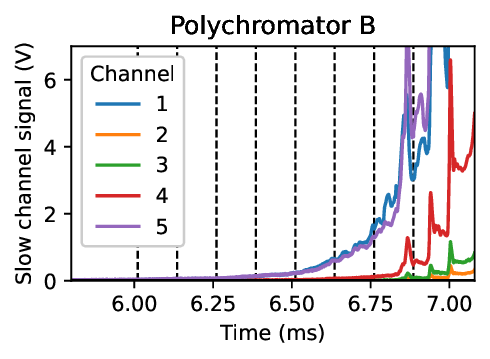}
    \caption{Polychromator slow channels. Thomson laser pulses are dashed vertical lines.}
    \label{fig:slow_channels}
\end{figure}

Polychromator slow channel data is shown in Figure~\ref{fig:slow_channels}. The slow channels recorded a rapid increase in the plasma background light on both A and B views around the time of the 5th laser pulse. The signal was strongest on channels A3 and B5, which were relatively more sensitive and had the widest spectral bandpass (see Figure~\ref{fig:poly_cals}). Background light was expected to increase due to the increase in plasma density during compression. Another contribution to the background increase might be light coming into view from plasma-surface interactions at the liner. The liner comes into view around $t=6.6~\textrm{ms}$ and $t=6.45~\textrm{ms}$ for views A and B respectively which roughly correspond with the start of the steeper rise in background light. 

\begin{figure}
    \centering
    \includegraphics[width=1\linewidth]{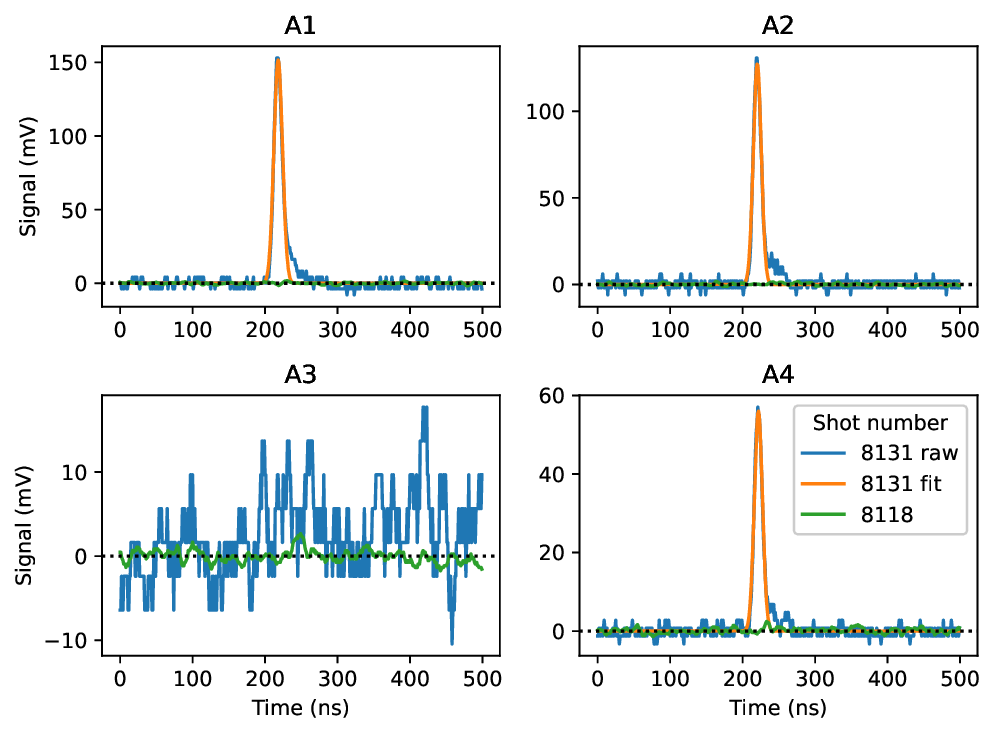}
    \includegraphics[width=1\linewidth]{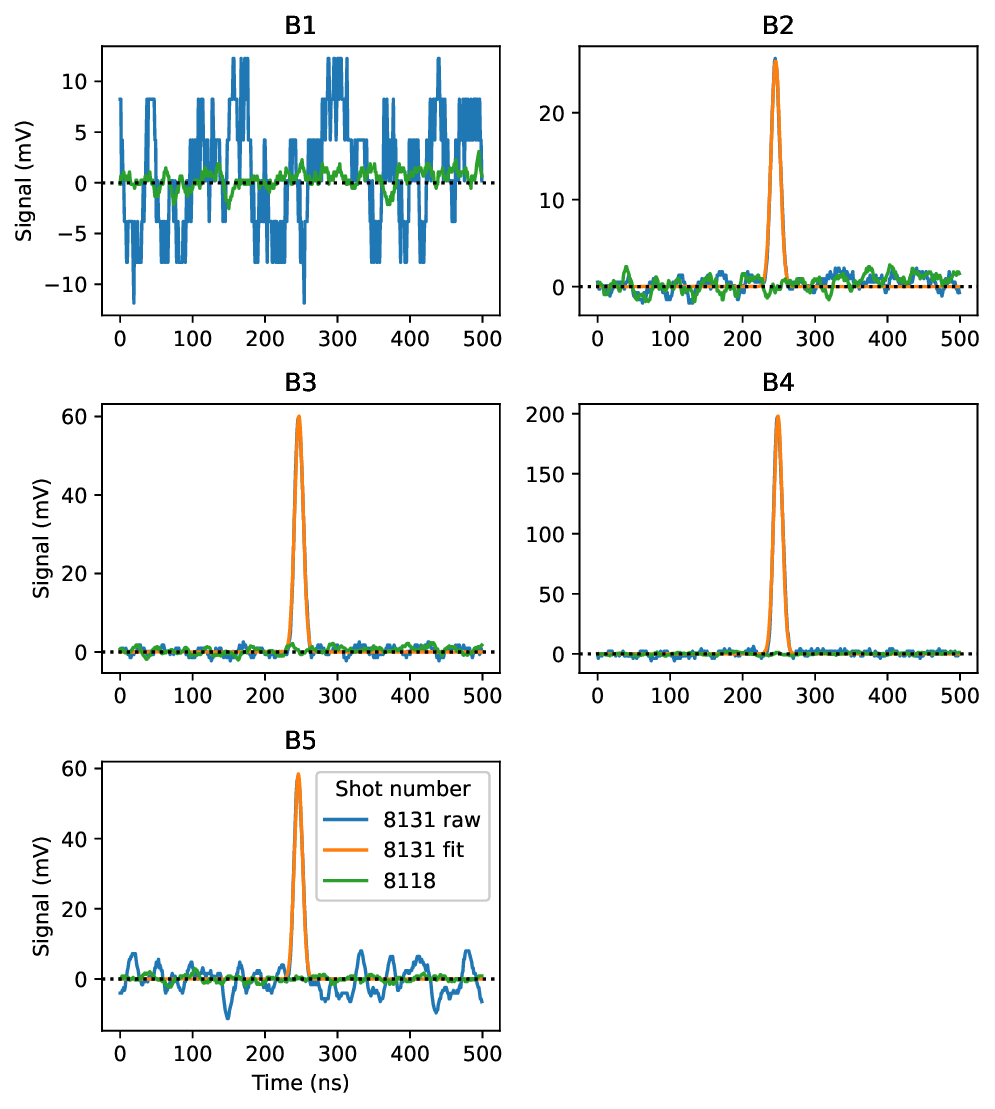}
    \caption{Polychromator fast channel raw data and Gaussian-fits for pulse 7 of LMC-12 (shot 8131). A vacuum test shot (8118) is also plotted to show that stray light was not significant. Larger noise in the raw signal for the plasma compression shot indicates the presence of stronger background plasma light.}
    \label{fig:fast_channels}
\end{figure}

Raw polychromator fast channel data is shown in Figure~\ref{fig:fast_channels} for laser pulse 7 on LMC-12. The total measured signal for both spatial points amounts to around 6000 photoelectron events (300-350~mV) which is roughly consistent with the expectations based on plasma density and the scattering geometry, and provides high confidence in the results. 

Vacuum shots, where the laser was fired without a plasma present, indicate that stray light was not significant leading up to LMC-12. Channels B1 (laser line stray light monitor) and A3 (broadband, short wavelength channel with no expected Thomson signal at these conditions) showed no change in stray light levels during the compression shot. 

\begin{figure}
    \centering
    \includegraphics[width=.7\linewidth]{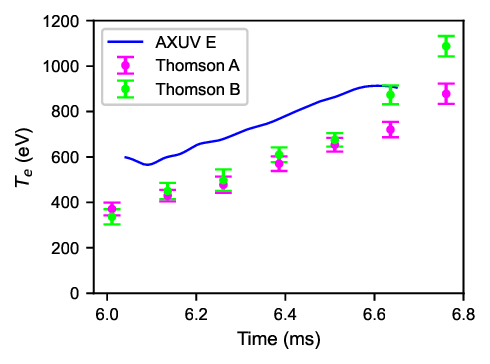}
    \caption{$T_e$ measured by Thomson scattering for pulses 1-7 on both views for LMC-12. Shown for comparison is $T_e$ from a multi-energy soft X-ray diagnostic (AXUV E). }
    \label{fig:temperatures}
\end{figure}

\begin{figure}
    \centering
    \includegraphics[width=.7\linewidth]{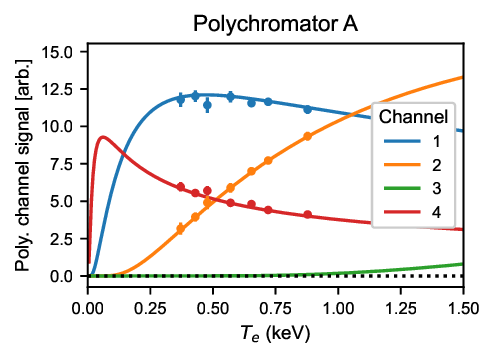}        \includegraphics[width=.7\linewidth]{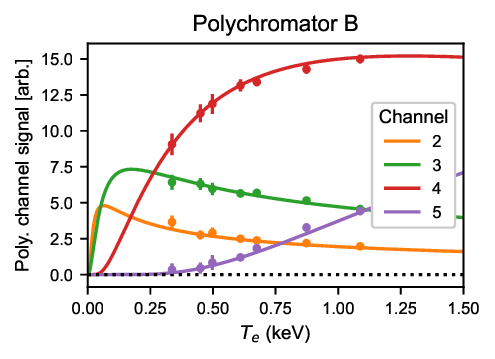}
    \caption{Polychromator channel measured (dots) and synthetic (lines) signals. Scaled measurements, $S_i/N$, and scaled error bars, $\sigma_{s,i}/N$, are shown for pulses 1 to 7. The measurements are plotted against the best fit $T_e$ and are in sequential order from left to right with pulse 1 corresponding to the leftmost set of points.}
    \label{fig:poly_fits}
\end{figure}

Figure~\ref{fig:temperatures} shows LMC-12 Thomson scattering $T_e$ measurements for laser pulses 1-7. Figure~\ref{fig:poly_fits} shows good fits between measured and synthetic data. For pulse 1 to pulse 5, views A and B report the same temperature which is consistent with the expectation that at earlier times the measurement points are roughly on the same flux surface. As the plasma core moves into view, a peaked temperature profile becomes more apparent since the two measurement points are measuring at increasingly distinct flux surfaces. The highest temperature measured is $T_e=1090\pm40~\rm{eV}$ for view B, pulse 7.

The measured rise in $T_e$ correlates with an independent measurement from a multi-energy soft X-ray diagnostic which infers $T_e$ from filtered AXUV measurements of the bremsstrahlung continuum \cite{howard2026science}. The AXUV diagnostic, which has more temporal information but larger measurement uncertainty, can use Thomson $T_e$ measurements for validation and normalization to compensate for systematic error \cite{delgado2007mcs}. 

\begin{figure}
    \centering
    \includegraphics[width=1\linewidth]{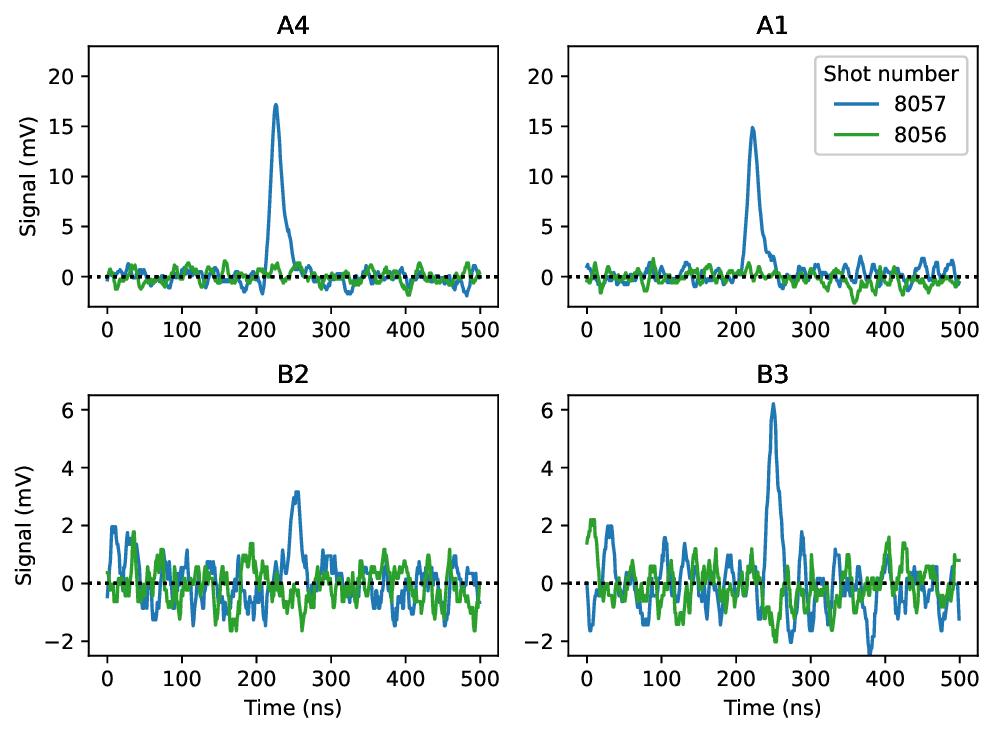}
    \caption{Polychromator fast channel data for an uncompressed plasma shot (8057) at $t=2.526~\rm{ms}$ after plasma formation. A vacuum test shot (8056) is provided to compare Thomson scattered light against stray light levels. The temperatures measured for views A and B were respectively $130 \pm 10~\rm{eV}$ and $ 100 \pm 30~\rm{eV}$.}
    \label{fig:plasma_shot}
\end{figure}

The system is sufficiently sensitive to measure Thomson scattering signal on uncompressed plasma shots where the density is on the order of 10 times lower. This is useful leading up to a compression shot to confirm that the collection optics are aligned to the laser. For an uncompressed plasma, the system is viewing a colder region well outside the plasma core, and so the Thomson signal is typically only observed on the first two spectral channels closest to the laser line as shown in Figure~\ref{fig:plasma_shot}. When the density is sufficiently high, a temperature can be estimated.

\section{Conclusion}

A Thomson scattering diagnostic was developed for LM26 and measured a peak electron temperature of $1090\pm40~\rm{eV}$ on the LMC-12 compression shot. The system is capable of measuring at 8 points in time and 2 points in space during compression. Access limitations constrained the design to a small scattering angle which significantly narrowed the spectral broadening of Thomson scattered light and provided some design advantages. Stray light was a negligible source of error. The design provides a strong basis for Thomson systems on future LM26 configurations where it will be required to measure core temperature pre-compression as well as deeper intro compression significantly past 1~keV.


\appendix

\section{Scattering Angle} \label{sec:scattering_angle_calculation}

For each view, the scattering angle is determined from 4 points in 3-dimensional (3D) space. The laser line is defined by the centres of the laser entrance window and the Brewster exit window. The viewing line is defined by the centre of the view window and by the spot projected onto the gun cavity wall (see Figure~\ref{fig:beam_path_side}) when backfilling the fiber bundle with white light. The coordinates of the first 3 reference points are obtained from a 3D model of the machine and their uncertainties are estimated from mechanical tolerances and by regularly monitoring the alignment of the laser and view with respect to the centre of these reference points.

\begin{figure}
    \centering
    \includegraphics[width=.9\linewidth]{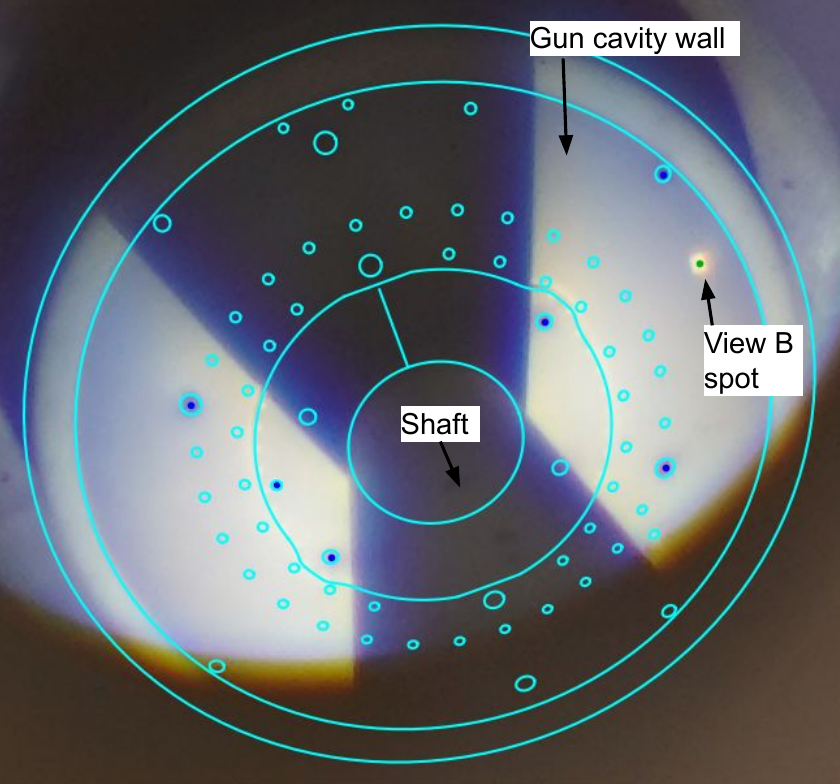}
    \caption{Image of the target chamber taken from a port on the access plate and looking towards the Marshall gun. The view B alignment light spot on the gun cavity wall is highlighted. The light blue shapes are features (e.g. port holes) from a 3D model that have been mapped onto the image. The dark blue dots are the features used to optimize the mapping.}
    \label{fig:reference_spot}
\end{figure}

The projected spot reference point was imaged with a camera a few hours before the compression shot (Figure~\ref{fig:reference_spot}). The coordinates of the spot are determined from an image analysis process which compares features in a 3D model to features in the camera image. A camera model is used to project real-world feature coordinates onto the image plane. An optimization is run to minimize the squared error between the positions of these feature sets. Once the optimal transformation is found, the position of the spot can be determined in real-world coordinates. The uncertainty in the spot's position was conservatively estimated to be $\pm 5~\rm{mm}$ from a 20~mm diameter spot.

Table~\ref{tab:reference_points} lists the reference coordinates and uncertainties used for the calculation of the scattering angles of the two views. Axial (Z) uncertainty is omitted. It is a small contribution to the overall uncertainty since the reference points defining the lines are long (metres) in the Z-direction relative to mechanical tolerances in Z (<20~mm).

\begin{table}[h]
\centering
\caption{Reference point coordinates. P1 is the centre of the view port. P2 is the centre of the fiber backfilled light spot on the gun cavity wall. P3 is the centre of the laser beam exit port. P4 is the centre of the laser beam entry port.}
\label{tab:reference_points}
\begin{tabular}{|l|c|c|c|}
\hline
 & \textbf{X (mm)} & \textbf{Y (mm)} & \textbf{Z (mm)} \\ \hline
P1A & 108 $\pm$ 4 & 63 $\pm$ 4 & 3499 \\ \hline
P1B & -100 $\pm$ 4 & 173 $\pm$ 4 & 3499 \\ \hline
P2A & 295 $\pm$ 5 & 176 $\pm$ 5 & 1742 \\ \hline
P2B & 379 $\pm$ 5 & 90 $\pm$ 5 & 1742 \\ \hline
P3  & 266 $\pm$ 4 & 154 $\pm$ 4 & 3615 \\ \hline
P4  & 0 $\pm$ 4   & 0 $\pm$ 4   & -1640 \\ \hline
\end{tabular}
\end{table}

\section{Polychromator Calibration} \label{sec:poly_calibration}
Polychromator spectral calibration was done in-house with a general calibration using a broadband pulsed light source, followed by a check and fine tuning of the calibration using a broadband DC light source.

\subsection{Pulsed Light Source}
An NKT Photonics SuperK Compact broadband light source was fed into a Horiba iHR320 monochromator using a 1200 grooves/mm grating. The monochromator was scanned through the wavelength range of interest from 990-1063~nm. A longpass filter (Thorlabs FELH0650) was added to the monochromator output to remove short wavelength harmonics. To measure the polychromator response, the monochrome light was fed to the polychromator via fiber. The pulse height on the relevant polychromator channel was measured on an oscilloscope for each test wavelength in the scan. In a separate scan over the same wavelength range, the monochrome light was sent to a Gentec PH100-Si-HA-D0 calibrated detector to measure the relative spectral power out of the monochromator. With these two datasets, the polychromator response was divided by the calibrated detector measurement to obtain the polychromator spectral calibration.

\subsection{DC Light Source}
A second calibration is done with a DC light source as a check and fine-tuning correction to the general calibration with the pulsed source. The same collection optics (lens, fiber) used for plasma measurements were setup to image light from a StellarNet UIS-LS DC light source. The distances between components were the same as for a plasma shot with the DC light source mimicking the plasma scattering volume. The collected light was delivered to the polychromator and the slow channel DC offsets were recorded. The "dark" offset with the light source off was subtracted from the signals with the light source on. The slow channel measurements were then compared to the expected signals knowing the light source spectral output provided by the manufacturer and the polychromator spectral calibration. The polychromator channel spectral calibrations were then scaled so the expected signals matched the slow channel measurements. The corrections for each spectral channel are shown in Table~\ref{tab:cal_correction_factors} and do not exceed $10\%$. The corrected calibrations only have a small effect on the reported temperature which is well within the error bars of the measurement.

\begin{table}[htbp]
    \centering
    \caption{Scalings to correct spectral calibrations of polychromators A and B based on the DC light source calibration.}
    \label{tab:cal_correction_factors}
    \begin{tabular}{|c|c|c|}
        \hline
        \textbf{\begin{tabular}[c]{@{}c@{}}Channel\end{tabular}} & \textbf{\begin{tabular}[c]{@{}c@{}}Poly. A \end{tabular}} & \textbf{\begin{tabular}[c]{@{}c@{}}Poly. B\end{tabular}} \\ \hline
        \textbf{1} & 1.00 & n/a \\ \hline
        \textbf{2} & 0.99 & 0.90 \\ \hline
        \textbf{3} & 1.00 & 1.09 \\ \hline
        \textbf{4} & 0.94 & 1.01 \\ \hline
        \textbf{5} & n/a & 1.00 \\ \hline
    \end{tabular}
\end{table}

\bibliography{refs}

\end{document}